\documentclass{sig-alternate}

\newcommand{\fmtResult}[1]{\pgfmathparse{#1}\pgfmathprintnumber[precision=3, assume math mode=true, zerofill]{\pgfmathresult}}

\newcommand\varStackexchangeLSIUPWarm{\fmtResult{0.635858782798}}
\newcommand\varStackexchangeLSILRWarm{\fmtResult{0.662366457988}}
\newcommand\varStackexchangeLightFMTagsAboutWarm{\textbf{\fmtResult{0.695144271163}}}
\newcommand\varStackexchangeLightFMTagsWarm{\fmtResult{0.674897071361}}
\newcommand\varStackexchangeLightFMTagsIdsWarm{\fmtResult{0.681874852145}}
\newcommand\varStackexchangeMFWarm{\fmtResult{0.541103233527}}
\newcommand\varStackexchangeLightFMTagsCold{\fmtResult{0.674506691956}}
\newcommand\varStackexchangeLSIUPCold{\fmtResult{0.637043290219}}

\newcommand\varStackexchangeLightFMTagsAboutCold{\textbf{\fmtResult{0.696337536662}}}
\newcommand\varStackexchangeLightFMTagsIdsCold{\fmtResult{0.674036514663}}
\newcommand\varStackexchangeMFCold{\fmtResult{0.508455758381}}
\newcommand\varStackexchangeLSILRCold{\fmtResult{0.659793439367}}

\newcommand\varMovielensLSIUPWarm{\fmtResult{0.687016462217}}
\newcommand\varMovielensLightFMTagsIdsWarm{\textbf{\fmtResult{0.762610614056}}}
\newcommand\varMovielensLSILRWarm{\fmtResult{0.686159232425}}
\newcommand\varMovielensMFWarm{\fmtResult{0.762156555535}}
\newcommand\varMovielensLightFMTagsWarm{\fmtResult{0.744316455133}}
\newcommand\varMovielensMFCold{\fmtResult{0.500360882628}}
\newcommand\varMovielensLightFMTagsIdsCold{\textbf{\fmtResult{0.715547376565}}}
\newcommand\varMovielensLSILRCold{\fmtResult{0.689780854354}}
\newcommand\varMovielensLightFMTagsCold{\fmtResult{0.707275991742}}
\newcommand\varMovielensLSIUPCold{\fmtResult{0.680680077337}}

\providecommand\varLightFMRepoLink{\url{https://github.com/lyst/lightfm/}}
\providecommand\varPaperRepoLink{\url{https://github.com/lyst/lightfm-paper/}}

\newcommand\symUserSet{U}
\newcommand\symItemSet{I}
\newcommand\symUserFeaturesSet{F^U}
\newcommand\symItemFeaturesSet{F^I}

\newcommand\symUserInteractionSet{S}

\providecommand\authAuthor{Maciej Kula}
\providecommand\authInstitution{Lyst}
\providecommand\authEmail{maciej.kula@lyst.com}

\usepackage{booktabs}
\usepackage{hyperref}
\usepackage{graphicx}
\usepackage{subcaption}
\usepackage{tabularx}
\usepackage{pgf}
\usepackage[input-decimal-markers={.}]{siunitx}

\begin{document}
%

\conferenceinfo{CBRecSys 2015,} {September 20, 2015, Vienna, Austria.}
\CopyrightYear{2015}
\newtoks\copyrightetc
\global\copyrightetc{Copyright remains with the authors and/or original copyright holders}

\title{Metadata Embeddings for User and Item Cold-start Recommendations}
%
%
%
%
%

\numberofauthors{1} 
%
\author{
%
%
\alignauthor
\authAuthor\\
       \affaddr{\authInstitution}\\
       \email{\authEmail}
}

\date{12 March 2015}

\maketitle
\begin{abstract}
I present a hybrid matrix factorisation model representing users and items as linear combinations of their content features' latent factors. The model outperforms both collaborative and content-based models in cold-start or sparse interaction data scenarios (using both user and item metadata), and performs at least as well as a pure collaborative matrix factorisation model where interaction data is abundant. Additionally, feature embeddings produced by the model encode semantic information in a way reminiscent of word embedding approaches, making them useful for a range of related tasks such as tag recommendations.
\end{abstract}

\category{H.3.3}{Information Storage and Retrieval}{Information Search and Retrieval}[Information Filtering]

\keywords{Recommender Systems, Cold-start, Matrix Factorization}

\section{Introduction}
Building recommender systems that perform well in cold-start scenarios (where little data is available on new users and items) remains a challenge. The standard matrix factorisation (MF) model performs poorly in that setting: it is difficult to effectively estimate user and item latent factors when collaborative interaction data is sparse.

Content-based (CB) methods address this by representing items through their metadata \cite{lops2011content}. As these are known in advance, recommendations can be computed even for new items for which no collaborative data has been gathered. Unfortunately, no transfer learning occurs in CB models: models for each user are estimated in isolation and do not benefit from data on other users. Consequently, CB models perform worse than MF models where collaborative information is available and require a large amount of data on each user, rendering them unsuitable for user cold-start \cite{adomavicius2005toward}.

At \authInstitution{}, solving these problems is crucial. We are a fashion company aiming to provide our users with a convenient and engaging way to browse---and shop---for fashion online. To that end we maintain a very large product catalogue: at the time of writing, we aggregate over 8 million fashion items from across the web, adding tens of thousands of new products every day.

Three factors conspire to make recommendations challenging for us. Firstly, our system contains a very large number of items. This makes our data very sparse. Secondly, we deal in fashion: often, the most relevant items are those from newly released collections, allowing us only a short window to gather data and provide effective recommendations. Finally, a large proportion of our users are first-time visitors: we would like to present them with compelling recommendations even with little data. This combination of user and item cold-start makes both pure collaborative and content-based methods unsuitable for us.

To solve this problem, I use a hybrid content-collaborative model, called LightFM due to its resemblance to factorisation machines (see Section \ref{sec:related}). In LightFM, like in a collaborative filtering model, users and items are represented as latent vectors (embeddings). However, just as in a CB model, these are entirely defined by functions (in this case, linear combinations) of embeddings of the content features that describe each product or user. For example, if the movie `Wizard of Oz' is described by the following features: \textit{`musical fantasy', `Judy Garland'}, and \textit{`Wizard of Oz'}, then its latent representation will be given by the sum of these features' latent representations. 

In doing so, LightFM unites the advantages of content-based and collaborative recommenders. In this paper, I formalise the model and present empirical results on two datasets, showing that:
\begin{enumerate}
\item In both cold-start and low density scenarios, LightFM performs at least as well as pure content-based models, substantially outperforming them when either \textit{(1)} collaborative information is available in the training set or \textit{(2)} user features are included in the model.
\item When collaborative data is abundant (warm-start, dense user-item matrix), LightFM performs at least as well as the MF model.
\item Embeddings produced by LightFM encode important semantic information about features, and can be used for related recommendation tasks such as tag recommendations.
\end{enumerate}
This has several benefits for real-world recommender systems. Because LightFM works well on both dense and sparse data, it obviates the need for building and maintaining multiple specialised machine learning models for each setting. Additionally, as it can use both user and item metadata, it has the quality of being applicable in both item and user cold-start scenarios.

To allow others to reproduce the results in this paper, I have released a Python implementation of LightFM\footnote{\varLightFMRepoLink}, and made the source code for this paper and all the experiments available on Github\footnote{\varPaperRepoLink}.

\section{LightFM}
\label{sec:model}
\subsection{Motivation}
The structure of the LightFM model is motivated by two considerations.
\begin{enumerate}
\item The model must be able to learn user and item representations from interaction data: if items described as `ball gown and `pencil skirt' are consistently all liked by users, the model must learn that ball gowns are similar to pencil skirts.
\item The model must be able to compute recommendations for new items and users.
\end{enumerate}
I fulfil the first requirement by using the latent representation approach. If ball gowns and pencil skirts are both liked by the same users, their embeddings will be close together; if ball gowns and biker jackets are never liked by the same users, their embeddings will be far apart.

Such representations allow transfer learning to occur. If the representations for ball gowns and pencil skirts are similar, we can confidently recommend ball gowns to a new user who has so far only interacted with pencil skirts.

This is over and above what pure CB models using dimensionality reduction techniques (such as latent semantic indexing, LSI) can achieve, as these only encode information given by feature co-occurrence rather than user actions. For example, suppose that all users who look at items described as aviators also look at items described as wayfarers, but the two features never describe the same item. In this case, the LSI vector for wayfarers will not be similar to the one for aviators even though collaborative information suggests it should be.

I fulfil the second requirement by representing items and users as linear combinations of their content features. Because content features are known the moment a user or item enters the system, this allows recommendations to be made straight away. The resulting structure is also easy to understand. The representation for denim jacket is simply a sum of the representation of denim and the representation of jacket; the representation for a female user from the US is a sum of the representations of US and female users.

\subsection{The Model}
To describe the model formally, let $\symUserSet$ be the set of users, $\symItemSet$ be the set of items, $\symUserFeaturesSet$ be the set of user features, and $\symItemFeaturesSet$ the set of item features. Each user interacts with a number of items, either in a favourable way (a positive interaction), or in an unfavourable way (a negative interaction). The set of all user-item interaction pairs $(u, i) \in \symUserSet \times \symItemSet$ is the union of both positive $\symUserInteractionSet^+$ and negative interactions $\symUserInteractionSet^-$. 

Users and items are fully described by their features. Each user $u$ is described by a set of features $f_u \subset \symUserFeaturesSet$. The same holds for each item $i$ whose features are given by $f_i \subset \symItemFeaturesSet$. The features are known in advance and represent user and item metadata.

The model is parameterised in terms of $d$-dimensional user and item feature embeddings $\boldsymbol{e}^U_f$ and $\boldsymbol{e}^I_f$ for each feature $f$. Each feature is also described by a scalar bias term ($b^U_f$ for user and $b^I_f$ for item features).

The latent representation of user $u$ is given by the sum of its features' latent vectors:
\begin{equation*}
\boldsymbol{q}_u = \sum_{j \in f_u}\boldsymbol{e}^U_j 
\end{equation*}
The same holds for item $i$:
\begin{equation*}
\boldsymbol{p}_i = \sum_{j \in f_i}\boldsymbol{e}^I_j
\end{equation*}
The bias term for user $u$ is given by the sum of the features' biases:
\begin{equation*}
b_u = \sum_{j \in f_u}b^U_j 
\end{equation*}
The same holds for item $i$:
\begin{equation*}
b_i = \sum_{j \in f_i}b^I_j
\end{equation*}

The model's prediction for user $u$ and item $i$ is then given by the dot product of user and item representations, adjusted by user and item feature biases:
\begin{equation}
\widehat{r}_{ui} = \mbox{\large $f$} \left(\boldsymbol{q}_u \cdot \boldsymbol{p}_i + b_u + b_i\right)
\end{equation}
There is a number of functions suitable for $f(\cdot)$. An identity function would work well for predicting ratings; in this paper, I am interested in predicting binary data, and so after Rendle \textit{et al.} \cite{rendle2009bpr} I choose the sigmoid function
\begin{equation*}
f(x) = \frac{1}{1 + \mathrm{exp}(-x)}.
\end{equation*}

The optimisation objective for the model consists in maximising the likelihood of the data conditional on the parameters. The likelihood is given by
\begin{equation}
L\left(\boldsymbol{e}^U, \boldsymbol{e}^I, \boldsymbol{b}^U, \boldsymbol{b}^I\right) = \prod_{(u, i) \in \symUserInteractionSet^+} \widehat{r}_{ui} \times \prod_{(u, i) \in \symUserInteractionSet^-} (1 - \widehat{r}_{ui})
\end{equation}

I train the model using asynchronous stochastic gradient descent \cite{recht2011hogwild}. I use four training threads for experiments performed in this paper. The per-parameter learning rate schedule is given by \textsc{Adagrad} \cite{duchi2011adaptive}.

\subsection{Relationship to Other Models}
The relationship between LightFM and the collaborative MF model is governed by the structure of the user and item feature sets. If the feature sets consist solely of indicator variables for each user and item, LightFM reduces to the standard MF model. If the feature sets also contain metadata features shared by more than one item or user, LightFM extends the MF model by letting the feature latent factors explain part of the structure of user interactions.

This is important on three counts. 
\begin{enumerate}
\item In most applications there will be fewer metadata features than there are users or items, either because an ontology with a fixed type/category structure is used, or because a fixed-size dictionary of most common terms is maintained when using raw textual features. This means that fewer parameters need to be estimated from limited training data, reducing the risk of overfitting and improving generalisation performance.
\item Latent vectors for indicator variables cannot be estimated for new, cold-start users or items. Representing these as combinations of metadata features that \emph{can} be estimated from the training set makes it possible to make cold-start predictions.
\item If only indicator features are present, LightFM should perform on par with the standard MF model.
\end{enumerate}

When only metadata features and no indicator variables are present, the model in general does not reduce to a pure content-based system. LightFM estimates feature embeddings by factorising the collaborative interaction matrix; this is unlike content-based systems which (when dimensionality reduction is used) factorise pure content co-occurrence matrices.

One special case where LightFM does reduce to a pure CB model is where each user is described by an indicator variable and has interacted only with one item. In that setting, the user vector is equivalent to a document vector in the LSI formulation, and only features which occur together in product descriptions will have similar embeddings.

The fact that LightFM contains both the pure CB model at the sparse data end of the spectrum and the MF model at the dense end suggests that it should adapt well to datasets of varying sparsity. In fact, empirical results show that it performs at least as well as the appropriate specialised model in each scenario.

\section{Related work}
\label{sec:related}
There are a number of related hybrid models attempting to solve the cold-start problem by jointly modelling content and collaborative data.

Soboroff \textit{et al.} \cite{soboroff1999combining} represent users as linear combinations of the feature vectors of items they have interacted with. They then perform LSI on the resulting item-feature matrix to obtain latent user profiles. Representations of new items are obtained by projecting them onto the latent feature space. The advantage of the model, relative to pure CB approaches, consists in using collaborative information encoded in the user-feature matrix. However, it models user preferences as being defined over individual features themselves instead of over items (sets of features). This is unlike LightFM, where a feature's effect in predicting an interaction is always taken in the context of all other features characterising a given user-item pair.

Saveski \textit{et al.} \cite{saveski2014item} perform joint factorisation of the user-item and item-feature matrices by using the same item latent feature matrix in both decompositions; the parameters are optimised by minimising a weighted sum of both matrices' reproduction loss functions. A weight hyperparameter governs the relative importance of accuracy in decomposing the collaborative and content matrices. A similar approach is used by McAuley \emph{et al.} \cite{mcauley2013hidden} for jointly modelling ratings and product reviews. Here, LightFM has the advantage of simplicity as its single optimisation objective is to factorise the user-item matrix.

Shmueli \textit{et al.} \cite{shmueli2012care} represent items as linear combinations of their features' latent factors to recommend news articles; like LightFM, they use a single-objective approach and minimise the user-item matrix reproduction loss. They show their approach to be successful in a modified cold-start setting, where both metadata and data on other users who have commented on a given article is available. However, their approach does not extend to modelling user features and does not provide evidence on model performance in warm-start scenario.

LightFM fits into the hybrid model tradition by jointly factorising the user-item, item-feature, and user-feature matrices. From a theory standpoint, it can be construed as a special case of Factorisation Machines \cite{rendle2010factorization}.

FMs provide an efficient method of estimating variable interaction terms in linear models under sparsity. Each variable is represented by a $k$-dimensional latent factor; the interaction between variable $i$ and $j$ is then given by the dot product of their latent factors. This has the advantage of reducing the number of parameters to be estimated.

LightFM further restricts the interaction structure by only estimating the interactions between user and item features. This aids the interpretability of resulting feature embeddings.

\section{Datasets}
\label{sec:datasets}
I evaluate LightFM's performance on two datasets. The datasets span the range of dense interaction data, where MF models can be expected to perform well (MovieLens), and sparse data, where CB models tend to perform better (CrossValidated). Both datasets are freely available.

\subsection{MovieLens}
The first experiment uses the well-known MovieLens 10M dataset\footnote{\url{http://grouplens.org/datasets/movielens/}}, combined with the Tag Genome tag set \cite{movielens:genome}.

The dataset consists of approximately 10 million movie ratings, submitted by $71,567$ users on $10,681$ movies. All movies are described by their genres and a list of tags from the Tag Genome. Each movie-tag pair is accompanied by a relevance score (between $0$ and $1$), denoting how accurately a given tag describes the movie.

To binarise the problem, I treat all ratings below $4.0$ (out of a 1 to 5 scale) as negative; all ratings equal to or above $4.0$ are positive. I also filter out all ratings that fall below the $0.8$ relevance threshold to retain only highly relevant tags.

The final dataset contains $69,878$ users, $10,681$ items, $9,996,948$ interactions, and $1030$ unique tags.

\subsection{CrossValidated}
The second dataset consists of questions and answers posted on CrossValidated\footnote{\url{http://stats.stackexchange.com}}, a part of the larger network of StackExchange collaborative Q\&A sites that focuses on statistics and machine learning. The dataset\footnote{\url{https://archive.org/details/stackexchange}} consists of $5953$ users, $44,200$ questions, and $188,865$ answers and comments. Each question is accompanied by one or more of $1032$ unique tags (such as `regression' or `hypothesis-testing'). Additionally, user metadata is available in the form of `About Me' sections on users' profiles.

The recommendation goal is to match users with questions they can answer. A user answering a question is taken as an implicit positive signal; all questions that a user has not answered are treated as implicit negative signals. For the training and test sets, I construct 3 negative training pairs for each positive user-question pair by randomly sampling from all questions that a given user has not answered.

To keep the model simple, I focus on a user's willingness to answer a question rather than their ability, and forego modelling user expertise \cite{san2014question}.

\section{Experimental setup}
\label{sec:setup}

For each dataset, I perform two experiments. The first simulates a warm-start setting: 20\% of all interaction pairs are randomly assigned to the test set, but all items and users are represented in the training set. The second is an item cold-start scenario: all interactions pertaining to 20\% of items are removed from the training set and added to the test set. This approximates a setting where the recommender is required to make recommendations from a pool of items for which no collaborative information has been gathered, and only content metadata (tags) are available.

I measure model accuracy using the mean receiver operating characteristics area under the curve (ROC AUC) metric. For an individual user, AUC corresponds to the probability that a randomly chosen positive item will be ranked higher than a randomly chosen negative item. A high AUC score is equivalent to low rank-inversion probability, where the recommender mistakenly ranks an unattractive item higher than an attractive item. I compute this metric for all users in the test set and average it for the final score.

I compute the AUC metric by repeatedly randomly splitting the dataset into a 80\% training set and a 20\% test set. The final score is given averaging across 10 repetitions.

I test the following models:
\begin{enumerate}
\item \textbf{MF}: a conventional matrix factorisation model with user and item biases and a sigmoid link function \cite{koren2009matrix}.
\item \textbf{LSI-LR}: a content-based model. To estimate it, I first derive latent topics from the item-feature matrix through latent semantic indexing and represent items as linear combinations of latent topics. I then fit a separate logistic regression (LR) model for each user in the topic mixture space. Unlike the LightFM model, which uses collaborative data to produce its latent representation, LSI-LR is purely based on factorising the content matrix. It should therefore be helpful in highlighting the benefit of using collaborative information for constructing feature embeddings.
\item \textbf{LSI-UP}: a hybrid model that represents user profiles (UP) as linear combinations of items' content vectors, then applies LSI to the resulting matrix to obtain latent user and item representations (\cite{soboroff1999combining}, see Section \ref{sec:related}). I estimate this model by first constructing a user-feature matrix: each row represents a user and is given by the sum of content feature vectors representing the items that user positively interacted with. I then apply truncated SVD to the normalised matrix to obtain user and feature latent vectors; item latent vectors are obtained through projecting them onto the latent feature space. The recommendations score for a user-item pair is then the inner product of their latent representations.
\item \textbf{LightFM (tags)}: the LightFM model using only tag features.
\item \textbf{LightFM (tags + ids)}: the LightFM model using both tag and item indicator features.
\item \textbf{LightFM (tags + about)}: the LightFM model using both item and user features. User features are available only for the CrossValidated dataset. I construct them by converting the `About Me' sections of users' profiles to a bag-of-words representation. I first strip them of all HTML tags and non-alphabetical characters, then convert the resulting string to lowercase and tokenise on spaces.
\end{enumerate}
In both LightFM (tags) and LightFM (tags + ids) users are described only by indicator features.

I train the LightFM models using stochastic gradient descent with an initial learning rate of 0.05. The latent dimensionality of the models is set to 64 for all models and experiments. This setting is intended to reflect the balance between model accuracy and the computational cost of larger vectors in production systems (additional results on model sensitivity to this parameter are presented in Section \ref{sec:sensitivity}). I regularise the model through an early-stopping criterion: the training is stopped when the model's performance on the test set stops improving.

\section{Experimental results}
\label{sec:experimental}

\subsection{Recommendation accuracy}
\begin{table}
\centering
\caption{Results}
\label{table:results}
\begin{tabular}{lcccc}
\toprule
& \multicolumn{2}{ c }{CrossValidated} & \multicolumn{2}{ c }{MovieLens} \\
\cmidrule(r){2-3}\cmidrule(r){4-5}
& {Warm} & {Cold} & {Warm} & {Cold} \\ \midrule
LSI-LR & \varStackexchangeLSILRWarm{} & \varStackexchangeLSILRCold{}  & \varMovielensLSILRWarm{} & \varMovielensLSILRCold{} \\
LSI-UP & \varStackexchangeLSIUPWarm{} & \varStackexchangeLSIUPCold{}  & \varMovielensLSIUPWarm{} & \varMovielensLSIUPCold{} \\
MF & \varStackexchangeMFWarm{} & \varStackexchangeMFCold{}  & \varMovielensMFWarm{} & \varMovielensMFCold{} \\
LightFM (tags) & \varStackexchangeLightFMTagsWarm{} & \varStackexchangeLightFMTagsCold{}  & \varMovielensLightFMTagsWarm{} & \varMovielensLightFMTagsCold{} \\
LightFM (tags + ids) & \varStackexchangeLightFMTagsIdsWarm{} & \varStackexchangeLightFMTagsIdsCold{}  & \varMovielensLightFMTagsIdsWarm{} & \varMovielensLightFMTagsIdsCold{} \\
LightFM (tags + about) & \varStackexchangeLightFMTagsAboutWarm{} & \varStackexchangeLightFMTagsAboutCold{}  & {} & {}\\

\bottomrule
\end{tabular}
\end{table}

Experimental results are summarised in Table \ref{table:results}. LightFM performs very well, outperforming or matching the specialised model for each scenario.

In the \textbf{warm-start, low-sparsity} case (warm-start MovieLens), LightFM outperforms MF slightly when using both tag and item indicator features. This suggest that using metadata features may be valuable even when abundant interaction data is present.

Notably, LightFM (tags) almost matches MF performance despite using only metadata features. The LSI-LR and LSI-UP models using the same information fare much worse. This demonstrates that \textit{(1)} it is crucial to use collaborative information when estimating content feature embeddings, and \textit{(2)} LightFM can capture that information much more accurately than other hybrid models such as LSI-UP.

In the \textbf{warm-start, high-sparsity} case (warm-start CrossValidated), MF performs very poorly. Because user interaction data is sparse (the CrossValidated user-item matrix is 99.95\% sparse vs only 99\% for the MovieLens dataset), MF is unable to learn good latent representations. Content-based models such as LSI-LR perform much better. 

LightFM variants provide the best performance. LightFM (tags + about) is by far the best model, showing the added advantage of LightFM's ability to integrate user metadata embeddings into the recommendation model. This is likely due to improved prediction performance for users with little data in the training set.

Results for the \textbf{cold-start} cases are broadly similar. On the CrossValidated dataset, all variants of LightFM outperform other models; LightFM (tags + about) again provides the best performance. Interestingly, LightFM (tags + indicators) outperforms LightFM (tags) slightly on the MovieLens dataset, even though no embeddings can be estimated for movies in the test set. This suggests that using both metadata and per-movie features allows the model to estimate better embeddings for both, much like the use of user and item bias terms allows better latent factors to be computed. Unsurprisingly, MF performs no better than random in the cold-start case.

In all scenarios the LSI-UP model performs no better than the LSI-LR model, despite its attempt to incorporate collaborative data. On the CrossValidated dataset it performs strictly worse. This might be because its latent representations are estimated on less data than in LSI-LR: as there are fewer users than items in the dataset, there are fewer rows in the user-feature matrix than in the item-feature matrix.

The results confirm that LightFM encompasses both the MF and the LSI-LR model as special cases, performing better than the LSI-LR model in the sparse-data scenario and better than the MF model in the dense-data case. This means not only that a single model can be maintained in either settings, but also that the model will continue to perform well even when the sparsity structure of that data changes.

Good performance of LightFM (tags) in both datasets is predicated on the availability of high-quality metadata. Nevertheless, it is often possible to obtain good quality metadata from item descriptions (genres, actor lists and so on), expert or community tagging (\emph{Pandora} \cite{westergren2007music}, \emph{StackOverflow}), or computer vision systems where image or audio data is available (we use image-based convolutional neural networks for product tagging). In fact, the feature embeddings produced by LightFM can themselves be used to assist the tagging process by suggesting related tags.

\subsection{Parameter Sensitivity}
\begin{figure*}
\caption{Latent dimension sensitivity}
\newcommand\figDim{3in}
\begin{subfigure}[b]{0.5\textwidth}
\centering
\includegraphics[height=\figDim, width=\figDim]{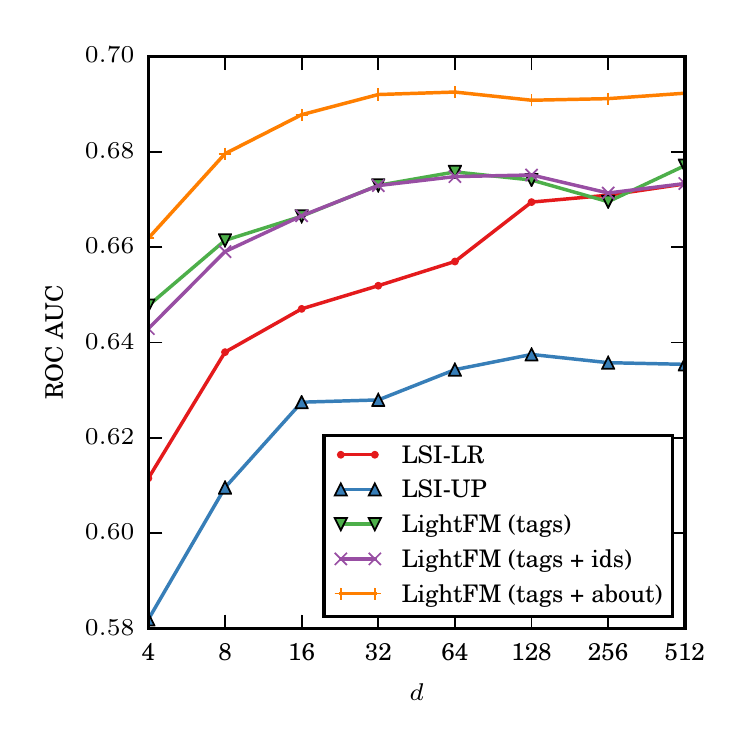}
\subcaption{CrossValidated}
\end{subfigure}
\begin{subfigure}[b]{0.5\textwidth}
\centering
\includegraphics[height=\figDim, width=\figDim]{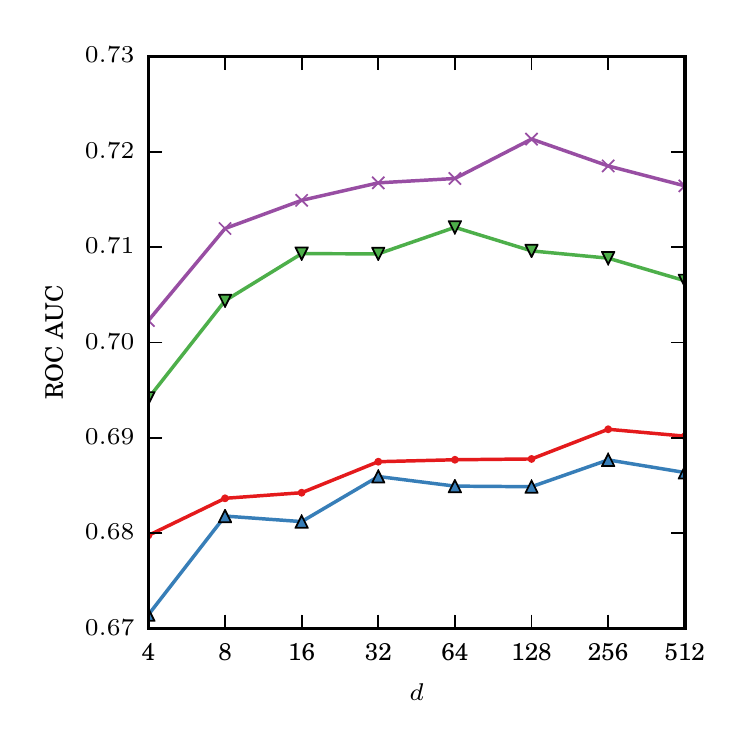}
\subcaption{MovieLens}
\end{subfigure}
\label{fig:power}
\end{figure*}
\label{sec:sensitivity}
Figure \ref{fig:power} plots the accuracy of LightFM, LSI-LR, and LSI-UP against values of the latent dimensionality hyperparameter $d$ in the cold-start scenario (averaged over 30 runs of each algorithm). As $d$ increases, each model is capable of modelling more complex structures and achieves better performance.

Interestingly, LightFM performs very well even with a small number of dimensions. In both datasets LightFM consistently outperforms other models, achieving high performance with as few as 16 dimensions. On CrossValidated data, it achieves the same performance as the LSI-LR model for much smaller $d$: it matches the accuracy of the 512-dimensional LSI-LR model even when using fewer than 32 dimensions.

This is an important win for large-scale recommender systems, where the choice of $d$ is governed by a trade-off between vector size and recommendation accuracy. Since smaller vectors occupy less memory and use fewer computations during query time, better representational power at small $d$ allows the system to achieve the same model performance at a smaller computational cost.

\subsection{Tag embeddings}
\begin{table}
\centering
\caption{Tag similarity}
\label{table:tags}
\begin{tabularx}{\columnwidth}{lX} \toprule
Query tag & Similar tags \\ \midrule
`regression' & `least squares', `multiple regression', `regression coefficients', `multicollinearity'\\
`MCMC' & `BUGS', `Metropolis-Hastings', `Beta-Binomial', `Gibbs', `Bayesian' \\
`survival' & `epidemiology', `Cox model', `Kaplan-Meier', `hazard' \\
`art house' & `pretentious', `boring', `graphic novel', `pointless', `weird' \\
`dystopia' & `post-apocalyptic', `futuristic', `artificial intelligence' \\
`bond' & `007', `secret service', `nuclear bomb', `spying', `assassin' \\
\bottomrule
\end{tabularx}
\end{table}

Feature embeddings generated by the LightFM model capture important information about the semantic relationships between different features. Table \ref{table:tags} gives some examples by listing groups of tags similar (in the cosine similarity sense) to a given query tag.

In this respect, LightFM is similar to recent word embedding approaches like word2vec and GloVe \cite{mikolov2013efficient, pennington2014glove}. This is perhaps unsurprising, given that word embedding techniques are closely related to forms of matrix factorisation \cite{levy2014neural}. Nevertheless, LightFM and word embeddings differ in one important respect: whilst word2vec and GloVe embeddings are driven by textual corpus co-incidence statistics, LightFM is based on user interaction data.

LightFM embeddings are useful for a number of recommendation tasks.
\begin{enumerate}
\item \textbf{Tag recommendation.} Various applications use collaborative tagging as a way of generating richer metadata for use in search and recommender system \cite{bastian2014linkedin, jaschke2007tag}. A tag recommender can enhance this process by either automatically applying matching tags, or generating suggested tags lists for approval by users. LightFM-produced tag embeddings will work well for this task without the need to build a separate specialised model for tag recommendations.
\item \textbf{Genre or category recommendation.} Many domains are characterised by an ontology of genres or categories which play an important role in the presentation of recommendations. For example, the Netflix interface is organised in genre rows; for \authInstitution{}, fashion designers, categories and subcategories are fundamental. The degree of similarity between the embeddings of genres or categories provides a ready basis for genre or category recommendations that respect the semantic structure of the ontology.
\item \textbf{Recommendation justification.} Rich information encoded in feature embeddings can help provide explanations for recommendations made by the system. For example, we might recommend a ball gown to a user who likes pencil skirts, and justify it by the two features' similarity as revealed by the distance between their latent factors.
\end{enumerate}

\section{Usage in Production Systems}
\label{sec:production}
The LightFM approach is motivated by our experience at \authInstitution{}. We have deployed LightFM in production, and successfully use it for a number of recommendation tasks. In this section, I describe some of the engineering and algorithm choices that make this possible.

\subsection{Model training and fold-in}
Thousands of new items and users appear on \authInstitution{} every day. To cope with this, we train our LightFM model in an online manner, continually updating the representations of existing features and creating fresh representations for features that we have never observed before.

We store model state, including feature embeddings and accumulated squared gradient information in a database. When new data on user interaction arrives, we restore the model state and resume training, folding in any newly observed features. Since our implementation uses per-parameter diminishing learning rates (\textsc{Adagrad}), any updates of established features will be incremental as the model adapts to new data. For new features, a high learning rate is used to allow useful embeddings to be learned as quickly as possible.

No re-training is necessary for folding in new products: their representation can be immediately computed as the sum of the representations of their features.

\vspace{10 mm}

\subsection{Feature engineering}
Each of our products is described by a set of textual features as well as structured metadata such as its type (dress, shoes and so on) or designer. These are accompanied by additional features coming from two sources.

Firstly, we employ a team of experienced fashion moderators, helping us to derive more fine-grained features such as clothing categories and subcategories (peplum dress, halterneck and so on).

Secondly, we use machine learning systems for automatic feature detection. The most important of these is a set of deep convolutional neural networks deriving feature tags from product image data.

\subsection{Approximate nearest neighbour searches}
The biggest application of LightFM-derived item representations are related product recommendations: given a product, we would like to recommend other highly relevant products. To do this efficiently across 8 million products, we use a combination of approximate (for on-demand recommendations) and exact (for near-line computation) nearest neighbour search.

For approximate nearest neighbour (ANN) queries, we use Random Projection (RP) trees \cite{dasgupta2008random, dasgupta2013randomized}. RP trees are a variant of random-projection \cite{dasgupta2000experiments} based locality sensitive hashing (LSH).

In LSH, $k$-bit hash codes for each point $\boldsymbol{x}$ are generated by drawing random hyperplanes $\boldsymbol{v}$, and then setting the $k$-th bit of the hash code to 1 if $\boldsymbol{x} \cdot \boldsymbol{v} \geq 0$ and 0 otherwise. The approximate nearest neighbours of $\boldsymbol{x}$ are then other points that share the same hash code (or whose hash codes are within some small Hamming distance of each other).

While extremely fast, LSH has the undesirable property of sometimes producing very highly unbalanced distribution of points across all hash codes: if points are densely concentrated, many codes of the tree will apply to no products while some will describe a very large number of points. This is unacceptable when building a production system, as it will lead to many queries being very slow.

RP trees provide much better guarantees about the size of leaf nodes: at each internal node, points are split based on the median distance to the chosen random hyperplane. This guarantees that at every split approximately half the points will be allocated to each leaf, making the distribution of points (and query performance) much more predictable.

\section{Conclusions and Future Work}
\label{sec:conclusions}
In this paper, I have presented an effective hybrid recommender model dubbed LightFM. I have shown the following:
\begin{enumerate}
\item LightFM performs at least as well as a specialised model across a wide range of collaborative data sparsity scenarios. It outperforms existing content-based and hybrid models in cold-start scenarios where collaborative data is abundant or where user metadata is available.
\item It produces high-quality content feature embeddings that capture important semantic information about the problem domain, and can be used for related tasks such as tag recommendations.
\end{enumerate}
Both properties make LightFM an attractive model, applicable both in cold- and warm-start settings. Nevertheless, I see two promising directions in extending the current approach.

Firstly, the model can be easily extended to use more sophisticated training methodologies. For example, an optimisation scheme using Weighted Approximate-Rank Pairwise loss \cite{weston2011wsabie} or directly optimising mean reciprocal rank could be used \cite{shi2012climf}.

Secondly, there is no easy way of incorporating visual or audio features in the present formulation of LightFM. At \authInstitution, we use a two-step process to address this: we first use convolutional neural networks (CNNs) on image data to generate binary tags for all products, and then use the tags for generating recommendations. We conjecture that substantial improvements could be realised if the CNNs were trained with recommendation loss directly.


%
\bibliographystyle{abbrv}
\bibliography{paper}  
%
%

\end{document}